\documentclass[RNAAS]{aastex62}

\begin{document}

\title{Multiwavelength Catalog of 10,000  4XMM-DR13 Sources with Known Classifications.  }


\correspondingauthor{Yichao Lin}
\email{yichaolin@gwu.edu}

\author{Yichao Lin}
\altaffiliation{}
\affiliation{The George Washington University}

\author{Hui  Yang}
\altaffiliation{}
\affiliation{The George Washington University}

\author{Jeremy Hare}
\altaffiliation{}
\affiliation{NASA Goddard Space Flight Center}
\affiliation{Center for Research and Exploration in Space Science and Technology}
\affiliation{The Catholic University of America}

\author{Igor Volkov}
\altaffiliation{}
\affiliation{The George Washington University}

\author{Oleg Kargaltsev}
\altaffiliation{}
\affiliation{The George Washington University}



\keywords{}



\section{Abstract}
We present a collection of $\sim10,000$ X-ray sources from the 4th XMM-Newton Serendipitous Source Catalog
 (4XMM-DR13) with literature-verified classifications and multi-wavelength (MW) counterparts. We describe the process 
by which MW properties are obtained and an interactive online visualization tool we developed.

\section{Introduction} 


Collections of reliably classified X-ray sources can be used for training supervised machine learning algorithms which can then quickly classify large numbers of X-ray sources  (see e.g., \citealt{2004ApJ...616.1284M,2022ApJ...941..104Y,2022A&A...657A.138T}). 
While bright X-ray sources can often be classified
solely 
from X-ray properties, the classification of 
 fainter but much more numerous
 sources 
 can greatly benefit  from including 
 properties of 
  their multiwavelength (MW) counterparts. 
 We created such dataset\footnote{\url{https://yichaolin-astro.github.io/4XMM-DR13-XCLASS/}} from 4XMM-DR13 \citep{2020A&A...641A.136W}. 


\section{Methods}

We searched the literature described in 
\cite{2021RNAAS...5..102Y,2022ApJ...941..104Y}, with the addition of the following class-specific catalogs  \cite{2020RNAAS...4..219J,2018ApJS..235....4O,2019ApJS..241...32L,2023A&A...671A.149F,2023A&A...677A.134N,2023A&A...675A.199A,2011AJ....142..181S,2009ApJ...696..870D,2023MNRAS.524.4867I,2015RAA....15.1095L}
, for reliably classified sources. 
These sources were sorted into 9 broad astrophysical classes:  active galactic nuclei (AGN), pulsars and isolated neutron stars (NS), non-accreting X-ray binaries (NS BIN)\footnote{Non-accreting binaries, including wide-orbit binaries with millisecond pulsars, red-back, and black widow systems \citep{2019ApJ...872...42S}.}, cataclysmic variables (CV), high-mass X-ray binaries (HMXB),  low-mass X-ray binaries (LMXB), high-mass stars (HM-STAR)\footnote{Includes Wolf-Rayet, O, B stars.}, low-mass stars (LM-STAR) and young stellar objects (YSO).  We then cross-matched these sources with 4XMM-DR13 within $r=10''$,
avoiding some particularly crowded environments (e.g.,  globular clusters, galaxies, the Galactic center), regions 
with complex 
diffuse X-ray emission (e.g., bright pulsar wind nebulae, supernova remnants) or 
infrared (IR)-bright fields (e.g., central parts of star-forming regions). 
Sources of populous classes (AGN, HM-STAR, LM-STAR, CV, and YSO) were 
omitted if 
the separations of their 4XMM-DR13 counterparts 
were larger than the combined  positional uncertainties (PUs) from the literature and 
 4XMM-DR13 
(at 95\% confidence) or $>3\arcsec$. For rare-type sources, we manually checked 
the counterparts by reviewing the 
publications on individual sources and inspecting the X-ray and MW images.  
Furthermore, all selected sources were matched to SIMBAD \citep{2000A&AS..143....9W}, and sources with classifications conflicting with the main SIMBAD class were omitted from the dataset (unless a mistake in the SIMBAD class 
was 
identified from looking at the original publications). 
For 8.5\% of our 4XMM-DR13
matches we were able to find matches in the Chandra Source Catalog version 2.1 (\citealt{2020AAS...23515405E}) which provides more accurate positional information that we then adopted for these sources.

As a next step, within $r=30''$ from each selected X-ray source, we combined sources from five all-sky lower-frequency  
 catalogs\footnote{
{\sl Gaia} DR3 \citep{2023A&A...674A...1G}, {\sl Gaia} EDR3 Distances \citep{2021AJ....161..147B},  Two Micron All-Sky Survey (2MASS; \citealt{2006AJ....131.1163S}), AllWISE \citep{2014yCat.2328....0C}, and CatWISE2020 \citep{2021ApJS..253....8M}} into 
a single merged MW catalog.
 In this process, the sources from  (near-)IR catalogs were first matched to Gaia DR3 sources, with Gaia source coordinates adjusted to each catalog's epoch using Gaia proper motions. If the separation between Gaia and other catalog sources 
 was
 smaller than the 5$\sigma$ 
 PUs of the two catalogs, similar to \cite{2019A&A...621A.144M}, the two sources were considered as a match. For (near-)IR sources 
 lacking Gaia DR3 counterparts, we 
 used CatWISE2020 as a reference catalog, instead of Gaia DR3. If no match with Gaia or CatWISE2020 is established, then the remaining sources from 2MASS and AllWISE catalogs 
 were matched with each other without any  
 proper motion corrections. We found that the 
 proper motion corrections 
 from CatWISE2020 are not as accurate as those from Gaia, and we did not use them if the ratios of the total proper motions to their uncertainties were  $<5$ or the reduced chi-squared of the astrometry fitting $>1.5$. We did not use the proper motions from AllWISE since they did not account for parallax. When determining the positions and the positional uncertainties of sources, we prioritize them from the Gaia, followed by CatWISE2020, 2MASS, and AllWISE catalogs, respectively, in cases where a source is matched to multiple catalogs. 
Finally, we 
used the probabilistic cross-matching algorithm NWAY \citep{2018MNRAS.473.4937S} to match the merged MW catalogs to the positions of X-ray sources while accounting for their individual positional uncertainties. Approximately 11\% of the X-ray sources have more than one MW counterpart matched within the cross-matching radius. These X-ray sources were omitted from further consideration to avoid confusion.

Additionally, extended sources with Gaia BP/RP flux excess factor phot\_bp\_rp\_excess\_factor$\footnote{A high value of phot\_bp\_rp\_excess\_factor indicates contamination in dense fields, or of extended objects \citep{2018A&A...616A...4E}.}>20$ or extended source flag$=5$ raised by AllWISE, and AGNs with large extinction values ($E(B-V)>0.05$) were removed from the TD. 
We also removed 
sources of HM-STARs, LM-STARs, and YSOs classes 
lacking MW counterparts because the sources from these classes are expected to have MW counterparts. 
Finally, we removed MW counterparts matched with isolated NSs by chance coincidence because 
virtually all 
 isolated NSs (except  Crab pulsar) are too faint to be detected in 
the surveys used here.

\begin{figure}[h!]
\begin{center}
\includegraphics[scale=0.33,angle=0]{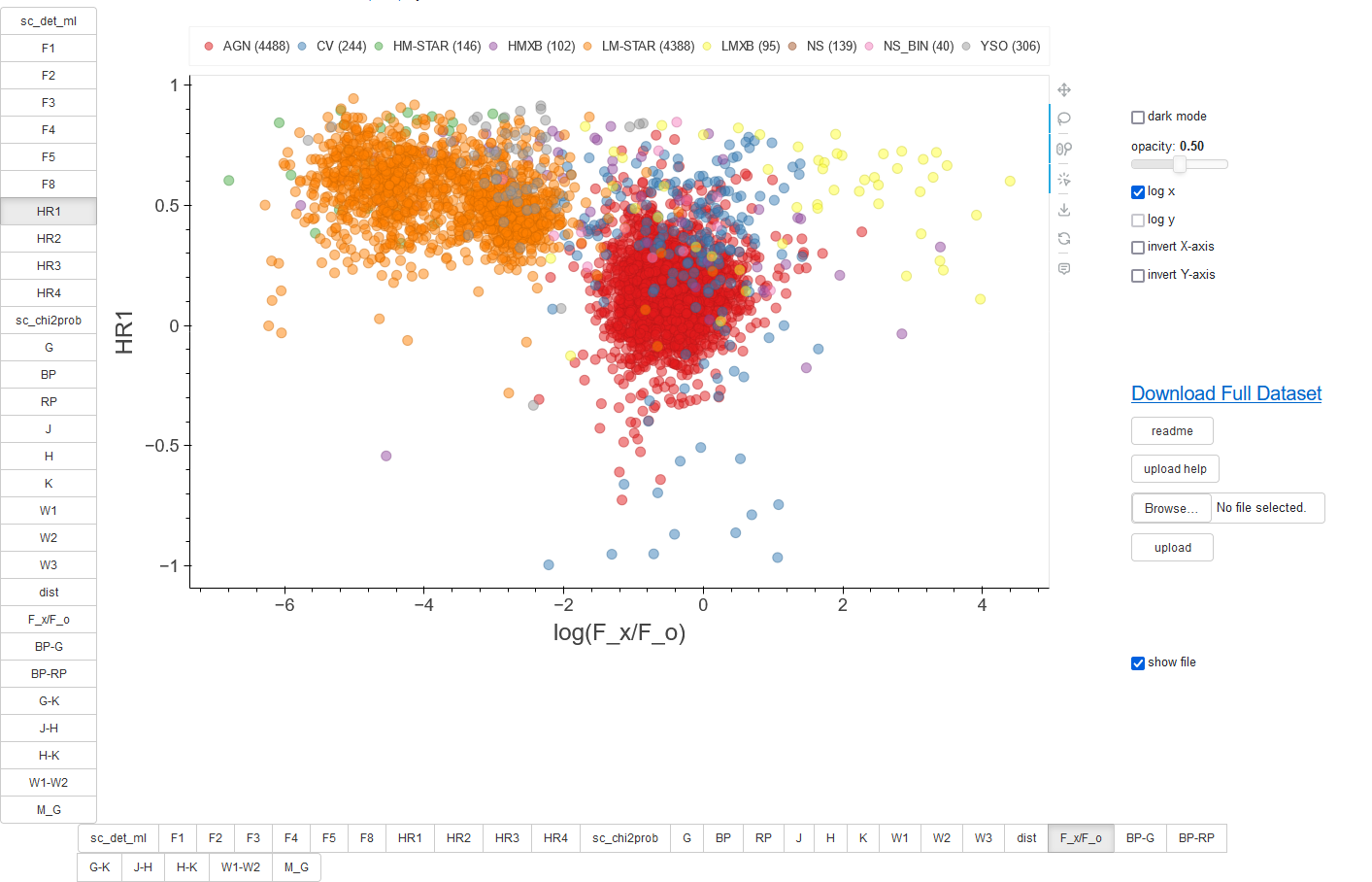}
\caption{A screenshot from the interactive visualization tool, available at \url{https://yichaolin-astro.github.io/4XMM-DR13-XCLASS/}, shows HR1 versus $F_{\rm X}/F_{\rm o}$, with an LM-STAR source selected. One can see LM-STAR and AGN are two separate clusters while other source classes show some overlapping, thus requiring using different pairs of source properties for distinction. 
\label{fig:1}}
\end{center}
\end{figure}

\section{THE VISUALIZATION TOOL GUI}

In total, we compiled 9,948 
 reliably classified 4XMM-DR13 sources into a comprehensive catalog which can be viewed  (see Figure \ref{fig:1}) 
  with our interactive visualization tool CIDView
\citep{2021RNAAS...5..102Y}.   The dataset is publicly available at  \url{https://yichaolin-astro.github.io/4XMM-DR13-XCLASS/}. 
The visualization tool allows for plotting of various permutations of two of the following features:
\begin{itemize}
\item 
X-ray properties\footnote{ described in detail in \url{http://xmmssc.irap.omp.eu/Catalogue/4XMM-DR13/col_unsrc.html}.} from 4XMM-DR13 include the energy fluxes in band 1 ($F1$; 0.2-0.5 keV), band 2 ($F2$; 0.5-1.0 keV), band 3 ($F3$; 1.0-2.0 keV), band 4 ($F4$; 2.0-4.5 keV), band 5 ($F5$; 4.5-12 keV), band 8 ($F8$; 0.2-12 keV), total band detection likelihood of the source (sc\_det\_ml), and probability of detection can be explained by a constant flux (sc\_chi2prob);
\item four X-ray hardness ratios derived from the energy fluxes using HR1 $= (F2-F1)/(F2+F1)$, HR2 $= (F3-F2)/(F3+F2)$, HR3 $= (F4-F3)/(F4+F3)$, HR4 $= (F5-F4)/(F5+F4)$;
\item $G$, ${G_{\rm BP}}$ (BP) and ${G_{\rm RP}}$ (RP) magnitudes from the Gaia eDR3;
\item $J$, $H$, and $K$ magnitudes from the 2MASS;
\item $W1$ and $W2$ magnitudes from the CatWISE2020 and $W3$ magnitude from the AllWISE;
\item a 
of optical and (near-)IR colors;
\item X-ray to optical flux ratio, $F_{\rm X}/F_{\rm o}$, with $F_{\rm X}=F8$ and $F_{\rm o}$ based on Gaia's $G$ band magnitude.
 \end{itemize} 
 Individual sources can also be selected for their information and MW images. 


\section{Acknowledgement}
This research has made use of data obtained from the 4XMM XMM-Newton serendipitous source catalogue compiled by the XMM-Newton Survey Science Centre consortium. This work was supported by NASA award 80NSSC22K1575. 

\end{document}